# Analytical solutions of symmetric isotropic spin clusters

Shadan Ghassemi Tabrizi*, Thomas D. Kühne

*Center for Advanced Systems Understanding (CASUS),
Am Untermarkt 20, 02826 Görlitz, Germany*
* s.ghassemi-tabrizi@hzdr.de

**Abstract.** Spin models like the Heisenberg Hamiltonian effectively describe the interactions of open-shell transition-metal ions on a lattice and can account for various properties of magnetic solids and molecules. Numerical methods are usually required to find exact or approximate eigenstates, but for small clusters with spatial symmetry, analytical solutions exist, and a few Heisenberg systems have been solved in closed form. This paper presents a simple, generally applicable approach to analytically solve isotropic spin clusters, based on adapting the basis to both total-spin and point-group symmetry to factor the Hamiltonian matrix into sufficiently small blocks. We demonstrate applications to small rings and polyhedra, some of which are straightforward to solve by successive spin-coupling for Heisenberg terms only; additional interactions, such as biquadratic exchange or multi-center terms necessitate symmetry adaptation.

## 1. Introduction

Spin Hamiltonians are used to model the properties of exchange-coupled magnetic solids and molecules [1], such as magnetic susceptibilities and heat capacities, or magnetic-resonance and neutron-scattering spectra. Based on matrix diagonalization, all physical quantities within the framework of the model can be calculated exactly. While the Hilbert space grows exponentially with the number of centers, the system-size limit for numerically exact or quasi-exact calculations [2] can be increased by taking advantage of symmetries [3–9]. However, for simulating large systems, various approximations are essential [10].

On the opposite side of the range, small clusters allow closed-form solutions, especially if the Hamiltonian is isotropic and invariant with respect to spin permutations [6] corresponding to spatial symmetries of the molecule or cluster. In addition to their pedagogical value, analytical solutions yield insights that might be obscured or unavailable in numerical data. For example, they can facilitate parametric plots of spectra and provide exact expressions for phase boundaries in parameter space to map precise quantum-phase diagrams, which would only be approximated in numerical computations.

Some Heisenberg systems are trivially integrable – solved without diagonalization or explicit adaptation to spatial symmetries – by Kambe's coupling method [11], which relies on successively forming subsystem spins in a hierarchical manner to ultimately produce total-spin multiplets. However, not all analytically solvable cases can be subjected to this approach, as it requires certain conditions on the coupling topology [12]. Besides, as explained in the Results section, Kambe's method ceases to be applicable when the model is extended to include additional isotropic terms, such as biquadratic exchange or multi-site interactions.

Here we find closed-form solutions for small isotropic clusters by exploiting spin and point-group (PG) symmetries to factorize the Hamiltonian into blocks, with each block corresponding to a specific irreducible representation (irrep) of SU(2) (quantum number $S$) and the point group (irrep label $\Gamma$). If the size of a block does not exceed $4\times 4$, then closed-form solutions for eigenvalues (energies) are guaranteed to exist, and have already been obtained for some Heisenberg clusters [13–16]. Specifically, for $s=\frac{1}{2}$ rings with $N=5,6,7$ sites [13,14], simultaneous adaptation to the $z$-component of spin (magnetization quantum number $M$) and the cyclic point group $C_N$ was sufficient to obtain subspaces of manageable sizes. For $s=\frac{1}{2}$, $N=8$ or $s=1$, $N=5$, adaptation to total spin ($S$ and $M$) and $C_N$ was achieved by a recursive technique designed for rings [15]. Finally, the ground state of the antiferromagnetic $s=\frac{1}{2}$ Heisenberg icosahedron was derived in a similar manner as presented here, but without explaining the method [16]. Our purpose is thus to provide a clear and easy-to-follow procedure to diagonalize any isotropic cluster that allows analytical solutions.

In the upcoming Theory section, we briefly revisit the symmetries of isotropic spin models and discuss various existing approaches for partitioning the Hamiltonian. We then explain our strategy of setting up a generalized (non-orthogonal) eigenvalue problem within a selected subspace by applying Löwdin's spin projector and a PG-projector to random states in an uncoupled basis. We also offer practical advice on implementing PG-symmetry. Our priority lies in providing a procedure that is as simple as possible to implement, rather than being the most computationally efficient, to enable the reader to calculate closed-form solutions for their cases of interest.

For small rings and polyhedra with various local spin values *s*, the Results section tabulates the dimensions of subspaces, provides selected energy expressions and boundary conditions in parameter space, and explores the ground state as a function of independent parameters. General isotropic spin Hamiltonians can involve a multitude of free parameters [17], making extensive tabulations of spectra or derived properties impractical. Our results are not directly intended to provide new insights into any exchange-coupled cluster but should assist in verifying independent implementations of the analytical diagonalization process, which could include additional terms.

## 2. Theory

**Symmetries of isotropic Hamiltonians.** In the Heisenberg model, pairwise interactions are parametrized by coupling constants $J_{ij}$, Eq. (1),

$$\hat{H}_J = \sum_{i<j} J_{ij} \hat{\mathbf{s}}_i \cdot \hat{\mathbf{s}}_j , \qquad (1)$$

where $\hat{\mathbf{s}}_i = (\hat{s}_{x,i}, \hat{s}_{y,i}, \hat{s}_{z,i})$ is the local spin vector of site *i*. Biquadratic exchange is another isotropic term, Eq. (2),

$$\hat{H}_K = \sum_{i<j} K_{ij} (\hat{\mathbf{s}}_i \cdot \hat{\mathbf{s}}_j)^2 . \qquad (2)$$

Note that $(\hat{\mathbf{s}}_i \cdot \hat{\mathbf{s}}_j)^2$ is a linear combination of scalar couplings of local spin operators of spherical tensor rank 1 (a Heisenberg-type contribution) and rank 2 [17]. The construction of rank-2 operators requires $s > \frac{1}{2}$, and therefore $\hat{H}_J$ is the only isotropic pairwise interaction for $s = \frac{1}{2}$. However, for $N \geq 4$, multi-center terms occur, see Results section.

Isotropy means invariance with respect to spin rotations [group SU(2)], due to commutation of the Hamiltonian with all components of the total spin $\hat{\mathbf{S}} = \sum_i \hat{\mathbf{s}}_i$, $[\hat{H}, \hat{S}_\alpha] = 0$ ($\alpha = x, y, z$). Each level of an isotropic $\hat{H}$ is a multiplet encompassing $2S+1$ states with *z*-projections ($\hat{S}_z$ eigenvalues) ranging from $M = -S$ to $M = +S$. All states of a single multiplet have an $\hat{\mathbf{S}}^2$ eigenvalue of $S(S+1)$. The basis can be spin-adapted by successive coupling, and the Hamiltonian matrix in a subspace with definite *S* is computed based on irreducible-tensor techniques, as explained in detail elsewhere [1,18]. In the frame of exact diagonalization of the Heisenberg model, the resulting reduction in matrix sizes is highly useful, and computational packages make such calculations accessible for studying various magnetic properties [19].

In addition to spin symmetry, the Heisenberg model, and indeed any isotropic spin model, is symmetric under permutations of sites according to the spatial symmetries of the cluster [6]. This spin-permutational symmetry (SPS) is often referred to as point-group (PG) symmetry. However, it is important to note that not all distinct PG symmetries of the electronic Hamiltonian are necessarily reflected in the isotropic spin model. For example, a planar hexanuclear cluster belonging to the molecular point group $D_{6h}$ could be represented as an $N = 6$ Heisenberg ring, with the latter model exhibiting only $D_6$ SPS. The full group $D_{6h} = D_6 \otimes C_i$ would pertain to an anisotropic spin Hamiltonian (not considered here) that more completely represents the physics by including the consequences of spin-orbit coupling [20]; some of the group operations would then represent combinations of spin permutations and spin rotations [21–23].

Combining total spin ($\hat{\mathbf{S}}^2$ and $\hat{S}_z$) with PG is significantly more complex than using either of these two symmetries separately. It is usually impossible to successively couple individual sites into larger subsystems and ultimately into a total-spin multiplet in a way that is compatible with the full point group, making demanding transformations between different coupling schemes unavoidable [6] (which are still manageable under specific circumstances [8,9]). Consequently, the application of PG symmetry is frequently limited either to a compatible subgroup [6] or – far more commonly –the full PG symmetry is utilized only in conjunction with $\hat{S}_z$ (instead of $\hat{\mathbf{S}}^2$ and $\hat{S}_z$) by working in an uncoupled basis $|m_1,...,m_N\rangle$ of definite local $z$-projections, $M = \sum_i m_i$ [6,7,24,25]. For a concise practical explanation of the latter strategy, see Ref. [7]. We briefly mention that an alternative technique for complete adjustment to full spin and PG symmetry relies on concepts from valence-bond theory but has not been widely adopted [26].

In contrast, we combine a PG-projection operator (see below) with Löwdin's projector [27] for full symmetry adaptation, with the aim of sufficiently reducing the dimensions of Hamiltonian blocks to enable analytical diagonalization. When used on a random state with definite $M$, Löwdin's projector, Eq. (3),

$$\hat{P}_S = \prod_{l \neq S} \frac{\hat{\mathbf{S}}^2 - l(l+1)}{S(S+1) - l(l+1)} \, , \tag{3}$$

affords a pure-spin state $|S,M\rangle$; all other contributions ($l \neq S$) are eliminated. A similar approach (also in conjunction with spatial symmetry) has occasionally been applied in numerical calculations, e.g., in Lanczos exact diagonalization for triangular-lattice cluster models [4] but was apparently not yet employed to obtain analytical solutions.

**Point-group projectors.** Here we will give a basic explanation of how to construct the PG-projection operators acting in spin space, where the point group is the set of SPS operations that leave the isotropic Hamiltonian invariant. These operations usually correspond to rotations or reflections of a geometrical shape that reflects the coupling topology, like a ring or a polyhedron, with spins located at the corners. For $N$ sites, the point group can be represented by permutation matrices $\mathbf{C}$ of size $N \times N$. To produce all $\mathbf{C}$, we rely on a set of generators from which all other group elements can be derived. In other words, every permutation matrix in the group is obtained either by repeated multiplication of the generators among themselves (e.g., a cyclic group $C_N$ of order $N$ has a single generator that corresponds to a rotation by $\frac{2\pi}{N}$) or through combinations and powers of several generators. The fundamental set of generators may not be unique, that is, different sets can generate the same group. For possible choices of the generating elements of the most relevant point groups, see, e.g., Ref. [28]. As an example, for the octahedral group $O_h$ with inversion (e.g., regular octahedron or cube), the minimal number of generators is two. These can be chosen as any $S_6$ axis with any $S_4$ (or $C_4$), where $S_6$ is a six-fold improper rotation axis and $C_4$ is a four-fold proper rotation axis. Alternatively, based on the group $O$ (without inversion, isomorphic to $T_d$, the group of the regular tetrahedron), which is generated by any $C_4$ with any $C_3$, one can add the inversion $C_i$ as a third generator to obtain the full group $O_h = O \times C_i$.

Through matrix multiplication, starting with the generators, new matrices are produced. This process is repeated until no new matrices are generated by any pairwise multiplications [29]. At this point, a faithful group representation in terms of $N$-dimensional permutation matrices $\mathbf{C}$ has been obtained. To derive all irreducible representations (irreps), we assign irrep matrices (listed, e.g., in the book by Herzig and Altman [30]) to each generator. When a new matrix $\mathbf{C}(k)$ emerges, $\mathbf{C}(i)\mathbf{C}(j) = \mathbf{C}(k)$, the irrep matrices are multiplied similarly, $\mathbf{D}^\Gamma(i)\mathbf{D}^\Gamma(j) = \mathbf{D}^\Gamma(k)$.

Finally, the $\mathbf{C}$ and $\mathbf{D}^\Gamma$ sets are used to construct the PG-projector $\hat{P}_\lambda^\Gamma$, defined in Eq. (4), for irrep $\Gamma$ and component $\lambda$ (the latter must be specified for multi-dimensional irreps, $d_\Gamma > 1$),

$$\hat{P}_\lambda^\Gamma = \frac{d_\Gamma}{h} \sum_{g=1}^{h} [D_{\lambda\lambda}^\Gamma(g)]^* \hat{G}(g) , \qquad (4)$$

where $h$ is the order of the group (the total number of elements $g$), $D_{\lambda\lambda}^{\Gamma}(g)$ is a diagonal entry of $\mathbf{D}^{\Gamma}(g)$, and $\hat{G}(g)$ is the respective symmetry operation in spin space; the asterisk (*) denotes complex conjugation.

As any permutation can be composed from pairwise exchanges, the exchange operator $\hat{P}_{ij}$ is needed to build $\hat{G}(g)$, cf. Figure 1. For $s = \frac{1}{2}$, $\hat{P}_{ij} = 1 + 4\hat{\mathbf{s}}_i \cdot \hat{\mathbf{s}}_j$ [31], but the exchange operator for arbitrary $s$ is less well known. Following Brown [32], $\hat{P}_{ij}$ is expanded in terms of powers of $\hat{\mathbf{s}}_i \cdot \hat{\mathbf{s}}_j$, Eq. (6).

$$\hat{P}_{ij} = \sum_{n=0}^{2s} A_n (\hat{\mathbf{s}}_i \cdot \hat{\mathbf{s}}_j)^n \quad (5)$$

The real coefficients $A_n$ (collected in vector $\mathbf{A}$) are derived from the linear system of Eq. (7),

$$\mathbf{MA} = \mathbf{v}, \quad (6)$$

where the vector $\mathbf{v}$ contains the exchange-parity (+1 or –1) of the coupled pair states with spin $s_{ij}$; starting with the symmetric ferromagnetic state ($s_{ij} = 2s$), the coupled levels are alternately symmetric (eigenvalue +1) and antisymmetric (–1) under exchange. The elements of matrix $\mathbf{M}$ are the respective powers of the eigenvalues of $\hat{\mathbf{s}}_i \cdot \hat{\mathbf{s}}_j = \frac{1}{2}[\hat{\mathbf{s}}_{ij}^2 - 2s(s+1)]$:

$$M(s_{ij}, n) = \left\{ \tfrac{1}{2}[s_{ij}(s_{ij}+1) - 2s(s+1)] \right\}^n . \quad (7)$$

It is then straightforward to construct the permutation operators $\hat{G}$ from the $N \times N$ permutation matrices $\mathbf{C}$ and the exchange operators $\hat{P}_{ij}$ by compounding pair exchanges.

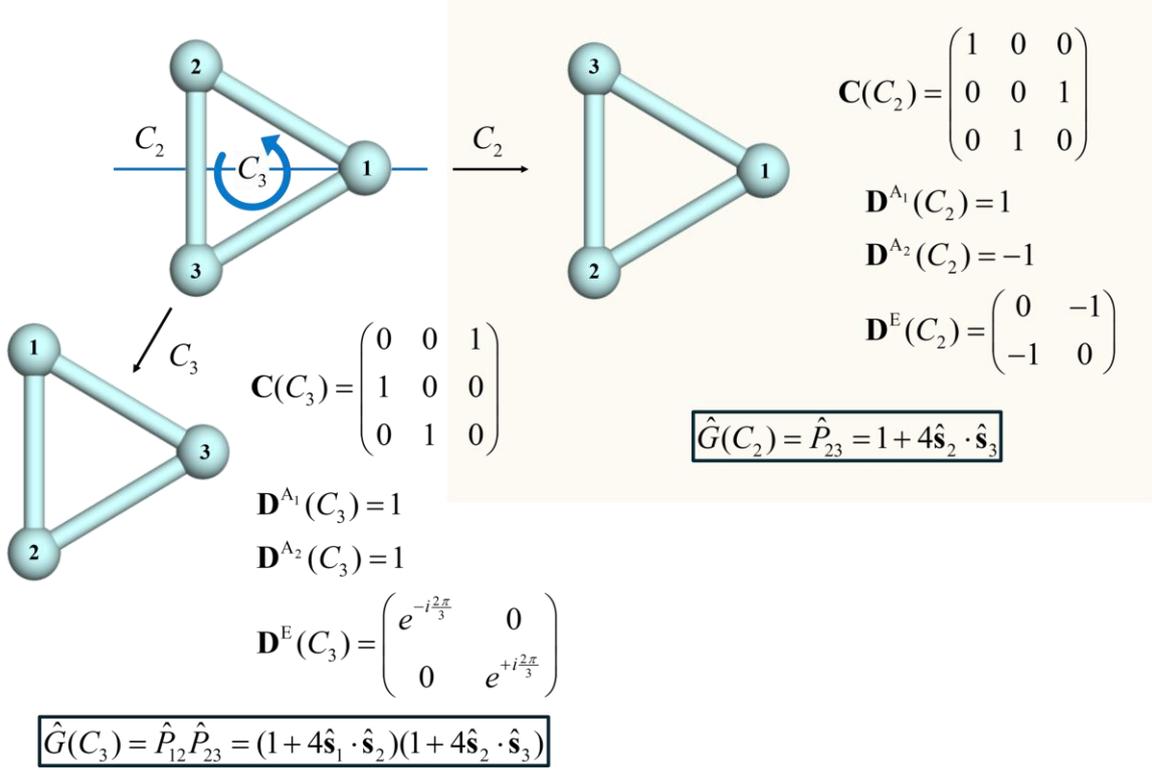

Figure 1: PG-symmetry in a triangular isotropic spin model (dihedral point group $D_3$). Two generators, a $C_3$ rotation and a $C_2$ rotation, are illustrated, alongside their permutation and irreducible representation (irrep) matrices. The expressions for exchange operators presented here are specific for $s = \frac{1}{2}$ sites.

**Generalized eigenvalue problem.** Symmetry projectors are idempotent, $\hat{P}^2 = \hat{P}$, and self-adjoint, $\hat{P}^\dagger = \hat{P}$, where the dagger ($^\dagger$) denotes the Hermitian adjoint (complex-conjugate transpose). Hence, their eigenvalues can only be 0 or 1. The dimension $d$ of the respective subspace is given by the trace, $d = \mathrm{Tr}(\hat{P})$. We calculate the Hamiltonian and overlap matrices, $\mathbf{h}$ and $\mathbf{s}$ (the latter is not to be mixed up with a spin vector), in a space $\mathbf{R} = (\mathbf{r}_1, ..., \mathbf{r}_d)$ of state vectors $\mathbf{r}_i$ comprising small random integers in a symbolic representation,

$$\mathbf{h} = \mathbf{R}^\dagger \mathbf{H} \mathbf{P}_S \mathbf{P}_\lambda^\Gamma \mathbf{R} , \tag{8}$$

$$\mathbf{s} = \mathbf{R}^\dagger \mathbf{P}_S \mathbf{P}_\lambda^\Gamma \mathbf{R} , \tag{9}$$

where $\mathbf{H}$, $\mathbf{P}_S$ and $\mathbf{P}_\lambda^\Gamma$ are the Hamiltonian, spin- and PG-projector representations, respectively, in the uncoupled basis $|m_1, ..., m_N\rangle$ for the selected magnetization $M$, and $d(S, M, \Gamma, \lambda) = \mathrm{Tr}(\mathbf{P}_S \mathbf{P}_\lambda^\Gamma)$. The generalized eigenvalue problem,

$$\mathbf{h}\mathbf{v} = E\mathbf{s}\mathbf{v} , \tag{10}$$

where **v** is an eigenvector, has real eigenvalues $E$ (energy levels). When solving Eq. (10) with symbolic computer algebra packages like `Mathematica` (`Eigenvalues[h,s]`) or the `MATLAB` symbolic toolbox (`eig(s\h)`), the energy expressions in spaces with $d = 3$ or $d = 4$ are usually very long, even when explicitly assuming symbolic Hamiltonian parameters to be real, and the built-in functions for algebraic simplification might not always produce significantly shorter forms. We observed that it is sometimes possible to obtain more concise results by a similarity transformation of **h**, Eq. (12),

$$\breve{\mathbf{h}} = \mathbf{L}^{-1}\mathbf{h}(\mathbf{L}^{-1})^{\dagger} \; , \tag{11}$$

based on the Cholesky decomposition, $\mathbf{s} = \mathbf{L}\mathbf{L}^{\dagger}$; $\breve{\mathbf{h}}$ has the same eigenvalues as the original problem, but, in contrast to $\mathbf{s}^{-1}\mathbf{h}$, it is Hermitian. In our experience, the symbolic eigenvalues of $\breve{\mathbf{h}}$ may be simpler than the corresponding equivalent expressions obtained from **h** and **s**. Still, most solutions of cubic ($d = 3$) or quartic ($d = 4$) polynomial equations are impractically lengthy functions of the parameters. Therefore, the Results section presents only a few illustrative and reasonably concise results for brevity.

**Additional considerations.** The only prerequisites for following our recipe are symbolic representations of the $\hat{\mathbf{s}}_i$ and the generators **C** for site permutations with their corresponding irrep matrices $\mathbf{D}^{\Gamma}$. Instead of constructing the $\hat{\mathbf{s}}_i$, one can directly compute the scalar products $\hat{\mathbf{s}}_i \cdot \hat{\mathbf{s}}_j$ of all relevant pairs in a magnetization subspace and build the projectors and all model terms considered in this paper (Heisenberg, biquadratic and four-center terms) from them.[1] This avoids working in the full Hilbert space throughout.

A symbolic calculation of $\mathbf{P}_S$ (with a significant fraction of non-zero elements) can become a bottleneck. Therefore, one may choose to first build the $(M,\Gamma,\lambda)$ basis. A direct full diagonalization of $\mathbf{P}_{\lambda}^{\Gamma}$, keeping only the eigenvectors with eigenvalue 1, is not always feasible with symbolic computer algebra, but is indeed not required, because the $(M,\Gamma,\lambda)$ space can be generated by scanning the rows of $\mathbf{P}_{\lambda}^{\Gamma}$ and selecting only the first column with a non-zero entry, discarding all other columns of $\mathbf{P}_{\lambda}^{\Gamma}$ that have a non-zero entry in the given row.[2] The idea

---

[1] Typically, $M = 0$ or $M = \frac{1}{2}$, which encompasses all multiplets.

[2] The described construction of a $(M,\Gamma,\lambda)$ subspace does not explicitly require the $\mathbf{P}_{\lambda}^{\Gamma}$ matrices and could instead be achieved as detailed in Ref. [7]. However, we believe that our current method, which applies (at least conceptually) a combined PG- and spin-projector to random states, offers more pedagogical clarity and would be

behind this procedure is that each uncoupled state $|m_1,...,m_N\rangle$ appears in at most one distinct state in the $(M,\Gamma,\lambda)$ basis.[3] The thus selected columns of $\mathbf{P}_\lambda^\Gamma$ are subsequently normalized and collected in a rectangular matrix $\mathbf{p}_\lambda^\Gamma$, which represents a complete orthonormal set in the $(M,\Gamma,\lambda)$ space, $(\mathbf{p}_\lambda^\Gamma)^\dagger \mathbf{p}_\lambda^\Gamma = \mathbf{1}$. Spin and Hamiltonian matrices are transformed accordingly, $\tilde{\mathbf{S}}^2 = (\mathbf{p}_\lambda^\Gamma)^\dagger \mathbf{S}^2 \mathbf{p}_\lambda^\Gamma$, $\tilde{\mathbf{H}} = (\mathbf{p}_\lambda^\Gamma)^\dagger \mathbf{H} \mathbf{p}_\lambda^\Gamma$, and $\tilde{\mathbf{P}}_S$ is constructed from $\tilde{\mathbf{S}}^2$ instead of $\mathbf{S}^2$. Finally, $\tilde{\mathbf{H}}$ and $\tilde{\mathbf{P}}_S$ are applied to a set of random states to set up the generalized eigenvalue problem in $(S,M,\Gamma,\lambda)$.

As long as $d(S,M,\Gamma,\lambda) \leq 4$ for all $S$, it may be possible to directly diagonalize $\tilde{\mathbf{H}}$, even when $d(M,\Gamma,\lambda) > 4$. On the other hand, if spin adaptation is necessary, an alternative to forming $\tilde{\mathbf{P}}_S$ is to diagonalize $\tilde{\mathbf{S}}^2$ and to then transform $\tilde{\mathbf{H}}$ into the space with the desired $S$. This method parallels how Schumann solved the Hubbard model on a square [33].[4] However, there is no guarantee that $\tilde{\mathbf{S}}^2$ can be diagonalized in symbolic form (at least not within a practical time frame), and this has indeed turned out to be impossible in some cases, like $M = 0$, $\Gamma = E_g$ ($d = 12$) in the $s = 1$ octahedron. Therefore, diagonalization of $\tilde{\mathbf{S}}^2$ is not a universal alternative to using Löwdin's projector.

## 3. Results

We focus on two rings (symmetric triangle and square) and three polyhedra (tetrahedron, octahedron and cube). Incidentally, except for the cube, these specific systems are trivially integrable within the Heisenberg model (see below), but they require matrix diagonalization when other isotropic terms are included. Tables of wave functions, energies, or other properties [21] are usually based on implicit assumptions about which independent parameters are

---

slightly simpler to implement. Forming $\mathbf{P}_\lambda^\Gamma$ in a symbolic representation usually does not pose a significant computational cost for systems that are small enough to have closed-form solutions.

[3] As noted in Ref. [22], this is not true for all multidimensional irreps in all point groups if one does not separate components $\lambda$ but instead uses a simplified projector, $\hat{P}^\Gamma = (d_\Gamma / h) \sum_g \chi^*(g) \hat{G}(g)$, based on characters, $\chi(g) = \text{Tr}[\mathbf{D}^\Gamma(g)]$, that summarily includes all components $\lambda$ of a multidimensional irrep $\Gamma$.

[4] Schumann additionally used so-called pseudospin symmetry, where applicable. This symmetry of bipartite Hubbard lattices does not exist in spin-only models. He adapted a basis with definite particle number and spin magnetization first to pseudospin, then to total spin, and lastly to PG-symmetry.

negligible. The systematic construction of all possible terms, whose number rises quickly with $N$ and $s$, was described in Ref. [17]. For instance, the isotropic Hamiltonian for a group of four $s = \frac{1}{2}$ sites has 9 independent parameters, whereas ten sites permit 8523 parameters [17] (this number would be lowered by spatial symmetries), although most of these would be negligibly small in practice. In our analysis, we primarily consider nearest-neighbor (NN) Heisenberg exchange and additionally consider biquadratic exchange or four-center terms. Our tables therefore do not aspire to be useful for analyzing all specific cases but should allow to effectively verify independent implementations by others. All energies are reported in units of the uniform NN coupling constant $J$, which is chosen to be antiferromagnetic, that is, we set $J = 1$.

**Triangle.** An $s = \frac{1}{2}$ triangle with three different coupling constants is the smallest system that necessitates matrix diagonalization, because there are two $S = \frac{1}{2}$ levels. On the other hand, an isosceles triangle has exchange symmetry, $[\hat{H}, \hat{P}_{12}] = 0$; with $J_{13} = J_{23}$, the square of the pair spin $\hat{\mathbf{s}}_{12} = \hat{\mathbf{s}}_1 + \hat{\mathbf{s}}_2$ is a good quantum number, $[\hat{H}_J, \hat{\mathbf{s}}_{12}^2] = 0$. This pair spin is then coupled with $\hat{\mathbf{s}}_3$ to obtain a total-spin multiplet,

$$\hat{H}_J = J_{12}\hat{\mathbf{s}}_1 \cdot \hat{\mathbf{s}}_2 + J_{13}(\hat{\mathbf{s}}_1 + \hat{\mathbf{s}}_2) \cdot \hat{\mathbf{s}}_3 = \\ \frac{J_{12}}{2}[\hat{\mathbf{s}}_{12}^2 - 2s(s+1)] + \frac{J_{13}}{2}[\hat{\mathbf{S}}^2 - \hat{\mathbf{s}}_{12}^2 - s(s+1)] \quad , \tag{12}$$

yielding the spectrum of Eq. (14):

$$E_J = \frac{J_{12}}{2}[s_{12}(s_{12}+1) - 2s(s+1)] + \frac{J_{13}}{2}[S(S+1) - s_{12}(s_{12}+1) - s(s+1)] \quad . \tag{13}$$

This is the simplest example of Kambe's method. In the symmetric triangle, all three couplings are equal, thus Eq. (12) becomes $\hat{H}_J = \frac{J}{2}[\hat{\mathbf{S}}^2 - 3s(s+1)]$, meaning that all multiplets with the same $S$ are degenerate [1].

However, when including biquadratic exchange in the isosceles triangle ($K_{13} = K_{23} \neq 0$), $\hat{P}_{12}$ remains a symmetry, $[\hat{H}_K, \hat{P}_{12}] = 0$, but $\hat{\mathbf{s}}_{12}^2$ is no longer a good quantum number, because $[(\hat{\mathbf{s}}_1 \cdot \hat{\mathbf{s}}_3)^2 + (\hat{\mathbf{s}}_2 \cdot \hat{\mathbf{s}}_3)^2, \hat{\mathbf{s}}_{12}^2] \neq 0$, rendering Kambe's method inapplicable.

Table 1 lists the subspace dimensions for symmetric triangles up to $s = \frac{13}{2}$, which is the smallest $s$ that does not permit obtaining the full spectrum in closed form, because there are five $(S = 6, \Gamma = E)$ levels. Note that Griffith had already classified terms in triangles, albeit for smaller $s$ [34]. Table 1 shows that up to $s = \frac{3}{2}$, there is at most one level in each $(S, \Gamma)$ sector, so symmetry adaptation suffices to determine the eigenfunctions, and all energies depend linearly on any parameters; phase boundaries in a two- or three-dimensional parameter space would then be straight lines or planes, respectively.

Table 1: Dimensions of combined spin- and PG-subspaces $(S, \Gamma)$ in symmetric triangles (point-group $D_3$) for various values $s$ (first column). Each cell represents the number of multiplets for ascending total-spin values $S$.[a]

| $s$ | $A_1$ | $A_2$ | $E$ |
|---|---|---|---|
| 1/2 | 0 1 0 | 0 0 0 | 1 0 0 |
| 1 | 0 1 0 1 | 1 0 0 0 | 0 1 1 0 |
| 3/2 | 0 1 1 0 1 | 0 1 0 0 0 | 1 1 1 1 0 |
| 2 | 1 0 1 1 1 0 1 | 0 1 0 1 0 0 0 | 0 1 2 1 1 1 0 |
| 5/2 | 0 1 1 1 1 1 0 1 | 0 1 1 0 1 0 0 0 | 1 1 2 2 1 1 1 0 |
| 3 | 0 1 0 2 1 1 1 1 0 1 | 1 0 1 1 1 0 1 0 0 0 | 0 1 2 2 2 2 1 1 1 0 |
| 7/2 | 0 1 1 1 2 1 1 1 1 0 1 | 0 1 1 1 1 1 0 1 0 0 0 | 1 1 2 3 2 2 2 1 1 1 0 |
| 4 | 1 0 1 1 2 1 2 1 1 1 1 0 1 | 0 1 0 2 1 1 1 1 0 1 0 0 0 | 0 1 2 2 3 3 2 2 2 1 1 1 0 |
| 9/2 | 0 1 1 1 2 2 1 2 1 1 1 1 0 1 | 0 1 1 1 2 1 1 1 1 0 1 0 0 0 | 1 1 2 3 3 3 3 2 2 2 1 1 1 0 |
| 5 | 0 1 0 2 1 2 2 2 1 2 1 1 1 1 0 1 | 1 0 1 1 2 1 2 1 1 1 1 0 1 0 0 0 | 0 1 2 2 3 4 3 3 3 2 2 2 1 1 1 0 |
| 11/2 | 0 1 1 1 2 2 2 2 2 1 2 1 1 1 1 0 1 | 0 1 1 1 2 2 1 2 1 1 1 1 0 1 0 0 0 | 1 1 2 3 3 4 4 3 3 3 2 2 2 1 1 1 0 |
| 6 | 1 0 1 1 2 1 3 2 2 2 2 1 2 1 1 1 1 0 1 | 0 1 0 2 1 2 2 2 1 2 1 1 1 1 0 1 0 0 | 0 1 2 2 3 4 4 4 4 3 3 3 2 2 2 1 1 0 |
| 13/2 | 0 1 1 1 2 2 2 3 2 2 2 2 1 2 1 1 1 1 0 1 | 0 1 1 1 2 2 2 2 2 1 2 1 1 1 1 0 1 0 0 0 | 1 1 2 3 3 4 5 4 4 4 3 3 3 2 2 2 1 1 1 0 |

[a]For example, for $s = \frac{1}{2}$, there are 0, 1 and 0 multiplets with PG-label $A_1$ for $S = 0$, 1 and 2, respectively.

For $s = 2$ and $s = \frac{5}{2}$, we exemplarily collect the full spectra as a function of the biquadratic-exchange constant $K$ in Table 2 and Table 3, respectively. (Our Hamiltonian is not exhaustive, e.g., three-center terms, which would occur for $s > \frac{1}{2}$, are ignored.)

Table 2: Full spectrum of an $s = 2$ triangle as a function of biquadratic exchange ($K$), with $J = 1$. The second column attributes numbers to all levels for easy comparison with Figure 2. Conditions on $K$ for a specific level to be the ground state are given in the last column.

| $(S, \Gamma)$ | # | Energy | Ground state |
|---|---|---|---|
| $(0, A_1)$ | 1 | $27K - \frac{9}{2}$ | $-\frac{1}{30} \leq K \leq \frac{1}{4}$ |
| $(1, A_2)$ | 2 | $40K - 4$ | $\varnothing$ |
| $(1, E)$ | 3 | $31K - 4$ | $\varnothing$ |
| $(2, A_1)$ | 4 | $72K - 3$ | $K \leq -\frac{1}{30}$ |

| (S,Γ) | # | Energy | Ground state |
|---|---|---|---|
| (2, E) | 5<br>6 | $\frac{93}{2}K \pm \frac{3}{2}\sqrt{193}\,|K|-3$ | ∅<br>∅ |
| (3, A$_1$) | 7 | $33K - \frac{3}{2}$ | ∅ |
| (3, A$_2$) | 8 | $15K - \frac{3}{2}$ | $K \geq \frac{1}{4}$ |
| (3, E) | 9 | $51K - \frac{3}{2}$ | ∅ |
| (4, A$_1$) | 10 | $37K + \frac{1}{2}$ | ∅ |
| (4, E) | 11 | $19K + \frac{1}{2}$ | ∅ |
| (5, E) | 12 | $24K + 3$ | ∅ |
| (6, A$_1$) | 13 | $48K + 6$ | ∅ |

Table 3: Energy spectrum and ground-state conditions for an $s = \frac{5}{2}$ triangle as a function of biquadratic exchange ($K$), with $J = 1$.

| (S,Γ) | # | Energy | Ground state |
|---|---|---|---|
| (1/2, E) | 1 | $\frac{1}{16}(975K - 102)$ | $-\frac{1}{46} \leq K \leq \frac{1}{9}$ |
| (3/2, A$_1$) | 2 | $\frac{1}{16}(867K - 90)$ | $\frac{1}{9} \leq K \leq \frac{7}{39}$ |
| (3/2, A$_2$) | 3 | $\frac{1}{16}(1155K - 90)$ | ∅ |
| (3/2, E) | 4 | $\frac{1}{16}(1443K - 90)$ | ∅ |
| (5/2, A$_1$) | 5 | $\frac{1}{16}(2447K - 70)$ | $K \leq -\frac{1}{46}$ |
| (5/2, A$_2$) | 6 | $\frac{1}{16}(1295K - 70)$ | ∅ |
| (5/2, E) | 7<br>8 | $\frac{1}{16}(-70 + 1679K \pm 24\sqrt{19}\,|K|)$ | ∅<br>∅ |
| (7/2, A$_1$) | 9 | $\frac{1}{16}(1971K - 42)$ | ∅ |
| (7/2, E) | 10<br>11 | $\frac{3}{16}(-14 + 401K \pm 32\sqrt{34}\,|K|)$ | ∅<br>∅ |
| (9/2, A$_1$) | 12 | $\frac{1}{16}(783K - 6)$ | ∅ |
| (9/2, A$_2$) | 13 | $\frac{1}{16}(399K - 6)$ | $K \geq \frac{7}{39}$ |
| (9/2, E) | 14 | $\frac{1}{16}(1359K - 6)$ | ∅ |
| (11/2, A$_1$) | 15 | $\frac{1}{16}(1091K + 38)$ | ∅ |
| (11/2, E) | 16 | $\frac{1}{16}(611K + 38)$ | ∅ |
| (13/2, E) | 17 | $\frac{1}{16}(975K + 90)$ | ∅ |
| (15/2, A$_1$) | 18 | $\frac{1}{16}(1875K + 150)$ | ∅ |

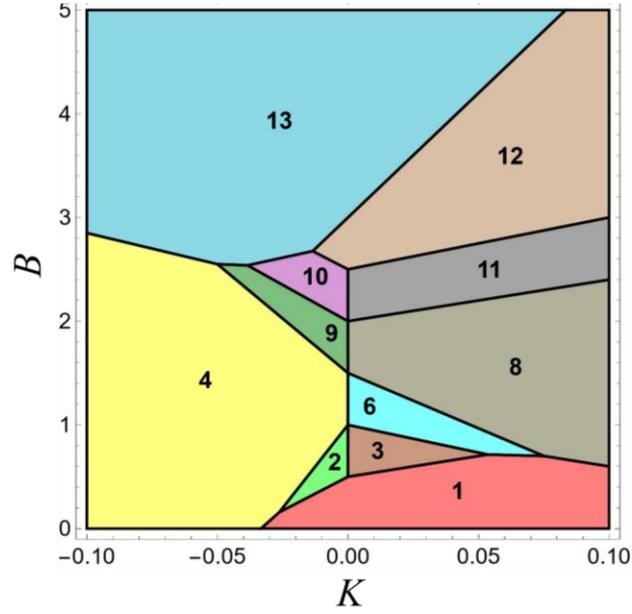

Figure 2: Diagram of the ground-state regions in a range of $K$, $B$ parameter space (see main text) of the antiferromagnetic $s = 2$ triangle $(J = 1)$. Each colored region corresponds to a distinct ground state (numbered according to the second column of Table 2), identified through analytical conditions derived from the spectrum (third column of Table 2), where a Zeeman energy-contribution of $BS$ was subtracted, because each respective ground state has magnetization $M = -S$ in a magnetic field applied along the $z$-axis.

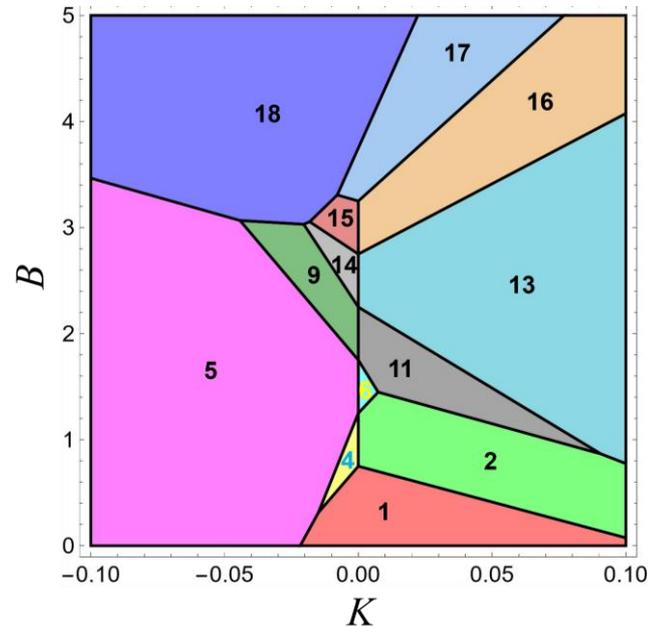

Figure 3: Diagram of the ground-state regions in a range of $K$, $B$ parameter space (see main text) of the antiferromagnetic $s = \frac{5}{2}$ triangle $(J = 1)$. For further details, see caption to Figure 2.

The ground states as a function of $K$ and magnetic-field strength $B$ (Zeeman term, $\hat{H}_B = B\hat{S}_z$) are plotted in Figure 2 ($s = 2$) and Figure 3 ($s = \frac{5}{2}$). These phase diagrams were built on simple analytical conditions, which resemble the ground-state conditions in the last column of Table 2 and Table 3, respectively, but also consider the magnetic field strength $B$ as another independent parameter.

**Square.** The Heisenberg square, with sites numbered consecutively, is again integrable by Kambe's method, because $[\hat{H}_J,\hat{s}_{13}^2]=[\hat{H}_J,\hat{s}_{24}^2]=0$, but $\hat{s}_{13}^2$ or $\hat{s}_{24}^2$ do not commute with the biquadratic $\hat{H}_K$. In addition, there exist two independent four-center interactions in the $s=\frac{1}{2}$ square:[5] the frequently discussed cyclic exchange, Eq. (15),

$$\hat{H}_C = C\left[(\hat{\mathbf{s}}_1 \cdot \hat{\mathbf{s}}_2)(\hat{\mathbf{s}}_3 \cdot \hat{\mathbf{s}}_4) + (\hat{\mathbf{s}}_1 \cdot \hat{\mathbf{s}}_4)(\hat{\mathbf{s}}_2 \cdot \hat{\mathbf{s}}_3) - (\hat{\mathbf{s}}_1 \cdot \hat{\mathbf{s}}_3)(\hat{\mathbf{s}}_2 \cdot \hat{\mathbf{s}}_4)\right] \quad (14)$$

and the less common non-cyclic exchange [17], Eq. (16),

$$\hat{H}_{\tilde{C}} = \tilde{C}\left[(\hat{\mathbf{s}}_1 \cdot \hat{\mathbf{s}}_2)(\hat{\mathbf{s}}_3 \cdot \hat{\mathbf{s}}_4) + (\hat{\mathbf{s}}_1 \cdot \hat{\mathbf{s}}_4)(\hat{\mathbf{s}}_2 \cdot \hat{\mathbf{s}}_3) + 6(\hat{\mathbf{s}}_1 \cdot \hat{\mathbf{s}}_3)(\hat{\mathbf{s}}_2 \cdot \hat{\mathbf{s}}_4)\right] \quad (15)$$

Except for $s=\frac{1}{2}$, $\hat{H}_C$ and $\hat{H}_{\tilde{C}}$ do not commute with $\hat{s}_{13}^2$ and $\hat{s}_{24}^2$. In other words, biquadratic exchange and multi-center interactions generally prevent trivial spin-coupling solutions. Dimensions of the $(S,\Gamma)$ sectors are listed in Table 4, and the spectrum for $s=1$ as a function of biquadratic and cyclic exchange ($K$ and $C$, respectively) is given in Table 5.

Table 4: Dimensions of the $(S,\Gamma)$ spaces in the square (point group $D_4$). For further information, see caption and footnote to Table 1.

| $s$ | $A_1$ | $A_2$ | $B_1$ | $B_2$ | E |
|---|---|---|---|---|---|
| 1 | 2 0 2 0 1 | 0 1 0 0 0 | 1 0 1 0 0 | 0 1 1 1 0 | 0 2 1 1 0 |
| 3/2 | 2 0 3 1 2 0 1 | 0 1 1 1 0 0 0 | 2 0 2 0 1 0 0 | 0 2 1 2 1 1 0 | 0 3 2 3 1 1 0 |
| 2 | 3 0 4 1 4 1 2 0 1 | 0 2 1 2 1 1 0 0 0 | 2 0 3 1 2 0 1 0 0 | 0 2 2 3 2 2 1 1 0 | 0 4 3 5 3 3 1 1 0 |
| 5/2 | 3 0 5 2 5 2 4 1 2 0 1 | 0 2 2 3 2 2 1 1 0 0 0 | 3 0 4 1 4 1 2 0 1 0 0 | 0 3 2 4 3 4 2 2 1 1 0 | 0 5 4 7 5 6 3 3 1 1 0 |

The ($K$, $C$) ground-state criteria are rather complicated and thus not shown here. However, one clear observation, which could not be obtained directly from numerical computations, is that, under the assumption of antiferromagnetic Heisenberg coupling, $J=1$, none of the levels **1**, **3**, **4**, **6**, **8**, **10**, **11**, **12**, **14**, or **15** is the ground state for any set ($K$, $C$). Phase diagrams including a magnetic field, setting either $K=0$ or $C=0$, are shown in Figure 4.

---

[5] For $s>\frac{1}{2}$, additional four-center (and three-center) interactions involve local rank-2 operators.

Table 5: Spectrum for an $s=1$ square as a function of biquadratic ($K$) and four-center coupling ($C$), at $J=1$. The ground-state conditions with $K$ set to zero are defined in the last column.

| $(S,\Gamma)$ | # | Energy | Ground state $(K=0)$ | Ground state $(C=0)$ |
|---|---|---|---|---|
| (0, A$_1$) | 1 | $8K + \frac{3}{2}C - \frac{3}{2} \pm \frac{1}{2}\sqrt{64K^2 + 96KC - 32K + 41C^2 - 34C + 9}$ | $\varnothing$ | $\varnothing$ |
|  | 2 |  | $C \leq \frac{1}{4}$ | $K \leq \frac{1}{2}$ |
| (0, B$_1$) | 3 | $8K + 3C - 1$ | $\varnothing$ | $\varnothing$ |
| (1, A$_2$) | 4 | $5K - 2C - \frac{1}{2}$ | $\varnothing$ | $\varnothing$ |
| (1, B$_2$) | 5 | $9K + 2C - \frac{5}{2}$ | $C = \frac{1}{4}$ | $\varnothing$ |
| (1, E) | 6 | $\frac{11}{2}K + C - \frac{3}{4} \pm \frac{1}{2}\sqrt{9K^2 + 20KC - K + 20C^2 - 10C + \frac{9}{4}}$ | $\varnothing$ | $\varnothing$ |
|  | 7 |  | $\varnothing$ | $K = \frac{1}{2}$ |
| (2, A$_1$) | 8 | $\frac{13}{2}K - \frac{3}{4} \pm \frac{1}{2}\sqrt{25K^2 - 48KC + K + 32C^2 + 8C + \frac{9}{4}}$ | $\varnothing$ | $\varnothing$ |
|  | 9 |  | $\frac{1}{4} \leq C \leq \frac{13}{2}$ | $K \geq \frac{1}{2}$ |
| (2, B$_1$) | 10 | $5K + \frac{1}{2}$ | $\varnothing$ | $\varnothing$ |
| (2, B$_2$) | 11 | $4K$ | $\varnothing$ | $\varnothing$ |
| (2, E) | 12 | $7K - 2C - \frac{1}{2}$ | $\varnothing$ | $\varnothing$ |
| (3, B$_2$) | 13 | $4K - 3C$ | $C \geq \frac{13}{2}$ | $\varnothing$ |
| (3, E) | 14 | $4K + C + 1$ | $\varnothing$ | $\varnothing$ |
| (4, A$_1$) | 15 | $4K + C + 2$ | $\varnothing$ | $\varnothing$ |

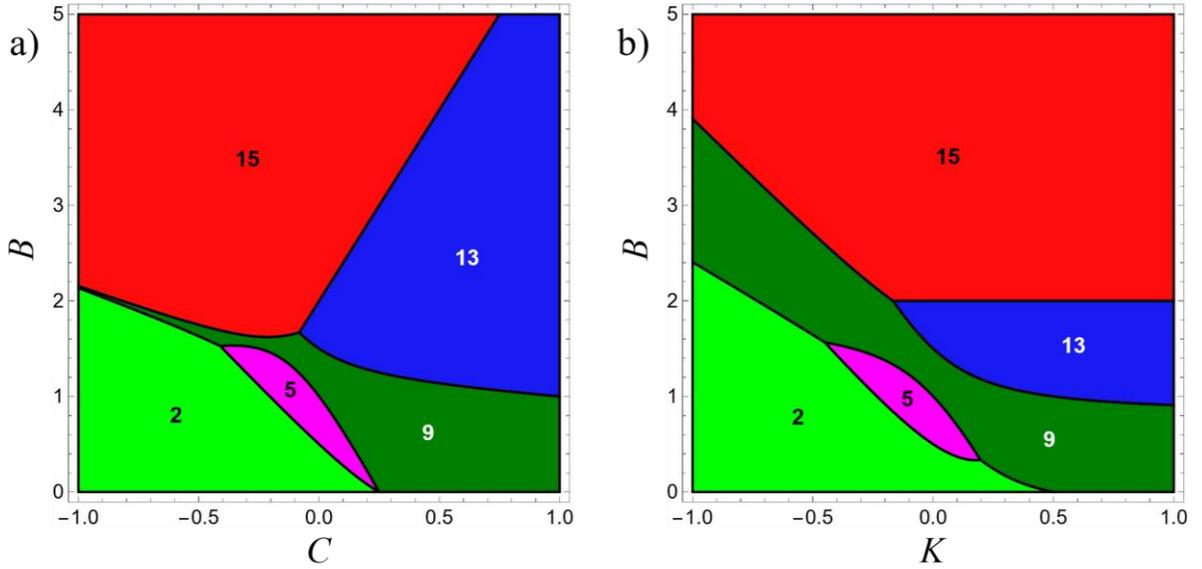

Figure 4: Ground-state regions for $K=0$ (plot a) and $C=0$ (plot b) in the $s=1$ square ($J=1$). For further details, see caption to Figure 2 and main text.

**Polyhedra.** Here we shall summarily discuss a few highly symmetric polyhedra. In the tetrahedron, all possible pairs are coupled equally, like in the triangle. Therefore, $\hat{H}_J$ is proportional to $\hat{\mathbf{S}}^2$, $\hat{H}_J = \frac{J}{2}[\hat{\mathbf{S}}^2 - 4s(s+1)]$. Biquadratic exchange again lifts degeneracies because it does not commute with pair spins, $[\hat{H}_K, \hat{s}_{12}^2] \neq 0$, etc. For $s = \frac{1}{2}$, where a single four-center term is compatible with $T_d$ symmetry, Eq. (17),

$$\hat{H}_T = T[(\hat{\mathbf{s}}_1 \cdot \hat{\mathbf{s}}_2)(\hat{\mathbf{s}}_3 \cdot \hat{\mathbf{s}}_4) + (\hat{\mathbf{s}}_1 \cdot \hat{\mathbf{s}}_4)(\hat{\mathbf{s}}_2 \cdot \hat{\mathbf{s}}_3) + (\hat{\mathbf{s}}_1 \cdot \hat{\mathbf{s}}_3)(\hat{\mathbf{s}}_2 \cdot \hat{\mathbf{s}}_4)] \ , \tag{16}$$

the pair spins commute with $\hat{H}_T$, but, as in the case of the square, the respective commutators are non-zero for $s > \frac{1}{2}$. Dimensions of symmetry subspaces are collected in Table 6. The dimensions for $s = 2$ were previously reported in Table 4 of Ref. [35].

Table 6: Dimensions of the $(S, \Gamma)$ spaces in the tetrahedron (point group $T_d$).

| $s$ | $A_1$ | $A_2$ | $E$ | $T_1$ | $T_2$ |
|---|---|---|---|---|---|
| 1/2 | 0 0 1 | 0 0 0 | 1 0 0 | 0 0 0 | 0 1 0 |
| 1 | 1 0 1 0 1 | 0 0 0 0 0 | 1 0 1 0 0 | 0 1 0 0 0 | 0 1 1 1 0 |
| 3/2 | 1 0 1 1 1 0 1 | 1 0 0 0 0 0 0 | 1 0 2 0 1 0 0 | 0 1 1 1 0 0 0 | 0 2 1 2 1 1 0 |
| 2 | 1 0 2 0 2 1 1 0 1 | 0 0 1 0 0 0 0 0 0 | 2 0 2 1 2 0 1 0 0 | 0 2 1 2 1 1 0 0 0 | 0 2 2 3 2 2 1 1 0 |
| 5/2 | 1 0 2 1 2 1 2 1 1 0 1 | 1 0 1 0 1 0 0 0 0 0 0 | 2 0 3 1 3 1 2 0 1 0 0 | 0 2 2 3 2 2 1 1 0 0 | 0 3 2 4 3 4 2 2 1 1 0 |
| 3 | 2 0 2 1 3 1 3 1 2 1 1 0 1 | 1 0 1 1 1 0 1 0 0 0 0 0 0 | 2 0 4 1 4 2 3 1 2 0 1 0 0 | 0 3 2 4 3 4 2 2 1 1 0 0 0 | 0 3 3 5 4 5 4 4 2 2 1 1 0 |

The complete analytical spectrum for the $s = \frac{3}{2}$ tetrahedron as a function of $K$ and $T$ is exemplarily provided in Table 7.

Table 7: Spectrum of the $s = \frac{3}{2}$ tetrahedron ($J = 1$) as a function of biquadratic and four-center coupling, $K$ and $T$.

| $(S,\Gamma)$ | # | Energy |
|---|---|---|
| $(0, A_1)$ | 1 | $\frac{327}{8}K + \frac{327}{16}T - \frac{15}{4}$ |
| $(0, A_2)$ | 2 | $\frac{135}{8}K + \frac{135}{16}T - \frac{15}{4}$ |
| $(0, E)$ | 3 | $\frac{423}{8}K + \frac{423}{16}T - \frac{15}{4}$ |
| $(1, T_1)$ | 4 | $\frac{179}{8}K + \frac{99}{16}T - \frac{13}{4}$ |
| $(1, T_2)$ | 5<br>6 | $\frac{299}{8}K + \frac{227}{16}T - \frac{13}{4} \pm \sqrt{81K^2 + 120KT + 46T^2}$ |
| (2, A$_1$) | 7 | $\frac{315}{8}K + \frac{267}{16}T - \frac{9}{4}$ |
| (2, E) | 8<br>9 | $\frac{243}{8}K + \frac{51}{16}T - \frac{9}{4} \pm \frac{3}{2}\sqrt{36K^2 + 36KT + 37T^2}$ |
| (2, T$_1$) | 10 | $\frac{219}{8}K + \frac{27}{16}T - \frac{9}{4}$ |
| (2, T$_2$) | 11 | $\frac{219}{8}K - \frac{69}{16}T - \frac{9}{4}$ |
| (3, A$_1$) | 12 | $\frac{303}{8}K - \frac{297}{16}T - \frac{3}{4}$ |
| (3, T$_1$) | 13 | $\frac{159}{8}K - \frac{81}{16}T - \frac{3}{4}$ |
| (3, T$_2$) | 14<br>15 | $\frac{279}{8}K - \frac{93}{16}T - \frac{3}{4} \pm \frac{3}{4}\sqrt{144K^2 - 360KT + 289T^2}$ |
| (4, A$_1$) | 16 | $\frac{287}{8}K - \frac{153}{16}T + \frac{5}{4}$ |
| (4, E) | 17 | $\frac{143}{8}K + \frac{63}{16}T + \frac{5}{4}$ |
| (4, T$_2$) | 18 | $\frac{191}{8}K - \frac{153}{16}T + \frac{5}{4}$ |
| (5, T$_2$) | 19 | $\frac{171}{8}K + \frac{27}{16}T + \frac{15}{4}$ |
| (6, A$_1$) | 20 | $\frac{243}{8}K + \frac{243}{16}T + \frac{27}{4}$ |

For most of the levels, the ($K$, $T$) criteria for a level to be the lowest-energy state are again quite complex. For $J = 1$, none of the levels **4**, **5**, **7**, **8**, **10**, **11**, **14**, **16**, **17**, **19**, or **20** can be the ground state for any ($K$, $T$), and **1** is a ground state only on the line $T = -2$, $K \geq -\frac{1}{28}$. Figure 5 shows phase diagrams including a magnetic field, setting $K = 0$ or $T = 0$.

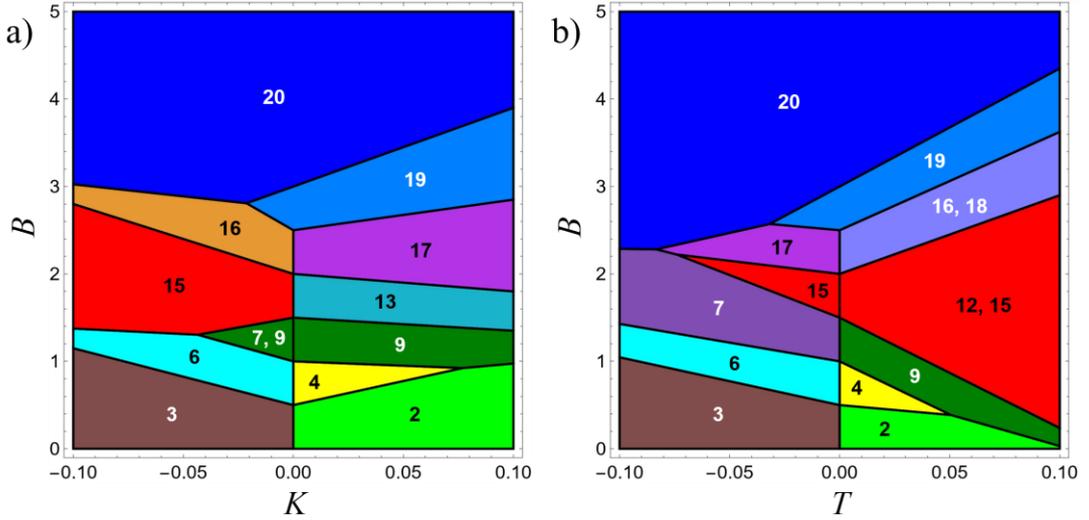

Figure 5: Ground-state regions for $T = 0$ (a) and $K = 0$ (b) in the $s = \frac{3}{2}$ tetrahedron ($J = 1$).

For the octahedron, a formulation of $\hat{H}_J$ as in Eq. (18),[59]

$$\hat{H}_J = \tfrac{J}{2}(\hat{\mathbf{S}}^2 - \hat{\mathbf{s}}_{14}^2 - \hat{\mathbf{s}}_{25}^2 - \hat{\mathbf{s}}_{36}^2) \;, \tag{17}$$

shows that eigenstates have definite pair spins for diametrically opposite sites. Accidental degeneracies between terms belonging to different $(S,\Gamma)$ sectors are lifted when incorporating multi-center terms [17] or biquadratic exchange. Table 8 shows that the $s = \tfrac{1}{2}$ octahedron is solved directly by spin- and PG-adaptation, and that the full spectrum can be obtained in closed form for $s = 1$ (not detailed here, due to excessively long expressions for solutions in three- and four-dimensional spaces).

Table 8: Dimensions of the $(S,\Gamma)$ spaces in the octahedron (point group $O_h$).

| $s$ | $A_{1g}$ | $A_{2g}$ | $A_{1u}$ | $A_{2u}$ | $E_g$ | $E_u$ | $T_{1g}$ | $T_{2g}$ | $T_{1u}$ | $T_{2u}$ |
|---|---|---|---|---|---|---|---|---|---|---|
| 1/2 | 0 1 0 1 | 1 0 0 0 | 0 0 0 0 | 1 0 0 0 | 0 1 1 0 | 0 0 0 0 | 0 0 0 0 | 0 1 0 0 | 1 0 1 0 | 0 1 0 0 |
| 1 | 3 0 3 1 2 0 1 | 0 2 0 2 0 0 0 | 1 0 0 0 0 0 0 | 0 1 0 1 0 0 0 | 1 2 4 2 2 1 0 | 0 1 1 0 0 0 0 | 0 2 1 1 0 0 0 | 2 1 3 1 1 0 0 | 0 4 2 3 1 1 0 | 1 2 3 2 1 0 0 |
| 3/2 | 0 5 2 7 3 4 2 2 0 1 | 3 1 4 3 4 1 2 0 0 0 | 0 2 0 2 0 0 0 0 0 0 | 3 0 3 1 2 0 1 0 0 0 | 1 6 8 8 7 6 3 2 1 0 | 1 2 4 2 2 1 0 0 0 0 | 2 4 6 5 4 2 1 0 0 0 | 0 7 5 8 4 4 1 1 0 0 | 5 4 11 7 9 4 4 1 1 0 | 1 7 7 9 6 5 2 1 0 0 |

Finally, the Heisenberg cube cannot be solved by Kambe's method, as it lacks conserved subsystem spins. The dimensions of the symmetry spaces for $s = \tfrac{1}{2}$, the only $s$ value for which the cube is fully solvable, are collected in Table 9. As far as we know, the complete spectrum of $\hat{H}_J$ was not derived previously, and we therefore present it in Table 10. We note accidental degeneracies for $E = -1, 0, +1, -1 \pm \sqrt{2}$.

Table 9: Dimensions of the $(S,\Gamma)$ spaces in the $s = \tfrac{1}{2}$ cube (point group $O_h$).

| $A_{1g}$ | $A_{2g}$ | $A_{1u}$ | $A_{2u}$ | $E_g$ | $E_u$ | $T_{1g}$ | $T_{2g}$ | $T_{1u}$ | $T_{2u}$ |
|---|---|---|---|---|---|---|---|---|---|
| 3 0 2 0 1 | 0 0 0 0 0 | 1 0 0 0 0 | 0 2 0 1 0 | 2 0 2 0 0 | 0 1 1 0 0 | 0 2 0 0 0 | 1 2 2 1 0 | 0 3 1 1 0 | 1 1 1 0 0 |

Table 10: Spectrum of the $s=\frac{1}{2}$ Heisenberg cube.

| $(S,\Gamma)$ | # | Energy |
|---|---|---|
| (0, A$_{1g}$) | 1<br>2<br>3 | $-\frac{5}{3}+\frac{2\sqrt{7}}{3}\cos\alpha \pm \frac{2\sqrt{21}}{3}\sin\alpha$<br>$-\frac{5}{3}-\frac{4\sqrt{7}}{3}\cos\alpha$ |
| (0, A$_{1u}$) | 4 | 0 |
| (0, E$_g$) | 5<br>6 | $-1\pm\sqrt{2}$ |
| (0, T$_{2g}$) | 7 | $-1$ |
| (0, T$_{2u}$) | 8 | $-2$ |
| (1, A$_{2u}$) | 9<br>10 | $-4$<br>0 |
| (1, E$_u$) | 11 | $-1$ |
| (1, T$_{1g}$) | 12<br>13 | $-\frac{1}{2}(1\pm\sqrt{5})$ |
| (1, T$_{2g}$) | 14<br>15 | $-\frac{1}{2}(3\pm\sqrt{5})$ |
| (1, T$_{1u}$) | 16<br>17<br>18 | $-\frac{2}{3}-\frac{\sqrt{10}}{3}\cos\beta \pm \frac{\sqrt{30}}{3}\sin\beta$<br>$-\frac{2}{3}+\frac{2\sqrt{10}}{3}\cos\beta$ |
| (1, T$_{2u}$) | 19 | 0 |
| (2, A$_{1g}$) | 20<br>21 | $-1\pm\sqrt{2}$ |
| (2, E$_g$) | 22<br>23 | $\frac{1}{2}(1\pm\sqrt{5})$ |
| (2, E$_u$) | 24 | 0 |
| (2, T$_{2g}$) | 25<br>26 | $\pm 1$ |
| (2, T$_{1u}$) | 27 | $-1$ |
| (2, T$_{2u}$) | 28 | 1 |
| (3, A$_{2u}$) | 29 | 0 |
| (3, T$_{2g}$) | 30 | 1 |
| (3, T$_{1u}$) | 31 | 2 |
| (4, A$_{1g}$) | 32 | 3 |

$\alpha = \frac{1}{3}\tan^{-1}\left(\frac{3\sqrt{591}}{13}\right)$, $\beta = \frac{1}{3}\tan^{-1}\left(3\sqrt{111}\right)$

## 4. Summary and Conclusion

While numerical methods are commonly used to investigate spin models of magnetic solids and molecules, small clusters with spatial symmetry allow for analytical diagonalization of the Hamiltonian, and a symbolic representation of spectra or wave functions may provide deeper insights that go beyond mere numerical data. To factor the Hamiltonian into symmetry subspaces and thus enable analytical diagonalization, we provided a simple yet effective approach to adapt the basis to both total-spin (using Löwdin's projector) and point-group (PG) symmetry. The construction of PG-projectors was extensively discussed. Overall, our procedure for employing spin- and PG-symmetry to set up a generalized eigenvalue problem in a subspace is not intended to be computationally optimal but designed to be easily followed and implemented. We chose small rings and polyhedra as examples and elucidated how additional interactions (beyond the Heisenberg model) prevent trivial integrability. Our aim was not to compile exhaustive tables but rather to highlight specific results that may be useful for verifying independent implementations of the analytical diagonalization scheme.

It is worth noting that a similar procedure would also be applicable to anisotropic systems. Although a general anisotropic Hamiltonian does not conserve spin, $[\hat{H},\hat{\mathbf{S}}^2] \neq 0$ and $[\hat{H},\hat{S}_z] \neq 0$, one can still make use of PG-symmetry and apply respective projectors to random states. With anisotropy, most site permutations must be combined with spin rotations to represent symmetries [21–23], and this necessitates working with the respective double group for systems with half-integer spin. Analytical solutions for anisotropic models are more severely restricted in terms of system size, because the group is smaller (lacking spin symmetry). Lastly, the present method could also be adapted for use with the Hubbard model, whose additional pseudospin symmetry on a bipartite lattice allows to further block-diagonalize the Hamiltonian [33,36]. However, the Hubbard model has itinerant electrons, and, at half-filling, its state space is larger than that of the $s = \frac{1}{2}$ Heisenberg model with the same number of sites. Therefore, the limits on system size are stricter.

Note that we leveraged symmetry to factor **H** into subspaces for easier diagonalization, but a classification of multiplets in terms of $S$ and $\Gamma$ holds qualitative value too, e.g., for deducing spectroscopic selection rules [20,37,38] or for assessing the momentum-transfer dependencies of inelastic neutron scattering intensities [20,38,39]. Symmetry classifications also aid in analyzing the mixing of multiplets by anisotropic terms [23,37].

Analytical solutions can be directly transferred when adding a further spin that has equal couplings to all other sites. The interaction of the original system with such a central site can be handled with Kambe's method, as detailed in Ref. [6]. However, the number of non-trivial systems that allow for closed-form solutions of their entire spectrum is naturally limited by the requirement that neither subspace dimension exceeds 4. In some situations, analytical diagonalization could still be used in the smaller spaces of large $S$ values, which are important to consider for ferromagnetic coupling or in high magnetic fields.